\documentclass[final]{IEEEtran}
\usepackage{amsmath}
\usepackage{amssymb}
\usepackage{xcolor}
\usepackage{graphicx}
\usepackage[noadjust]{cite}
\usepackage{balance}
\usepackage{booktabs}

\title{Deep Unfolding for Communications Systems: \\A Survey and Some New Directions}

\author{\IEEEauthorblockN{Alexios Balatsoukas-Stimming$^{1,2}$ and Christoph Studer$^3$} \\[0.1cm]
\IEEEauthorblockA{\em\small $^1$Telecommunications Circuits Laboratory, {\'E}cole polytechnique f{\'e}d{\'e}rale de Lausanne, Lausanne, Switzerland}\\
\IEEEauthorblockA{$^2$Electronic Systems Group, Eindhoven University of Technology, Eindhoven, The Netherlands (a.k.balatsoukas.stimming@tue.nl)}\\
\IEEEauthorblockA{$^3$School of Electrical and Computer Engineering, Cornell University, Ithaca, USA (studer@cornell.edu)}\\
\thanks{The work of ABS was supported by the Swiss NSF project PZ00P2\_179686. The work of CS was supported in part by Xilinx Inc.\ and the US NSF under grants ECCS-1408006, CCF-1535897, CCF-1652065,
CNS-1717559, and ECCS-1824379. The authors would like to thank O.~Casta\~neda~\oscapulpo{} and T.~Goldstein for discussions on deep unfolding.}
} 

\newcommand{\Ab}{\mathbf{A}}
\newcommand{\Cb}{\mathbf{C}}
\newcommand{\Ib}{\mathbf{I}}
\newcommand{\Hb}{\mathbf{H}}
\newcommand{\Pb}{\mathbf{P}}
\newcommand{\Wb}{\mathbf{W}}

\newcommand{\bb}{\mathbf{b}}

\newcommand{\kb}{\mathbf{k}}
\newcommand{\nb}{\mathbf{n}}
\newcommand{\rb}{\mathbf{r}}

\newcommand{\vb}{\mathbf{v}}
\newcommand{\yb}{\mathbf{y}}
\newcommand{\xb}{\mathbf{x}}
\newcommand{\zb}{\mathbf{z}}

\DeclareRobustCommand{\oscapulpo}{%
  \includegraphics[width=0.29cm]{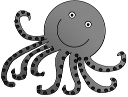}%
}

\begin{document}
\bstctlcite{IEEEexample:BSTcontrol}
\maketitle
\pagenumbering{gobble}
\begin{abstract}
Deep unfolding is a method of growing popularity that fuses iterative optimization algorithms with tools from neural networks to efficiently solve a range of tasks in machine learning, signal and image processing, and communication systems. 
This survey summarizes the principle of deep unfolding and discusses its recent use for communication systems with focus on detection and precoding in multi-antenna (MIMO) wireless systems and belief propagation decoding of error-correcting codes. 
To showcase the efficacy and generality of deep unfolding, we describe a range of other tasks relevant to communication systems that can be solved using this emerging paradigm.
We conclude the survey by outlining a list of open research problems and future research directions.  
\end{abstract}


\section{Introduction}

A large number of signal processing tasks in communications systems, such as detection and decoding, can be formulated as optimization problems. 
These optimization problems are typically solved using numerical algorithms that iteratively refine the solution.
Most practical communications applications require solving of these problems at high throughput and low latency, which implies that one can afford only a very small number of algorithm iterations (e.g., ten or fewer). 
In order to find accurate solutions with a small number of iterations, numerical solvers require careful parameter tuning (e.g., step-size selection).  
While the numerical optimization literature has focused extensively on analyzing convergence rates and stability given step-size conditions~\cite{Bertsekas15}, only very little is know about optimal parameter tuning under stringent iteration constraints. 
In practice, the algorithm parameters are typically set using heuristics (e.g., tuned by hand using simulations) or pessimistic bounds (e.g., given by the Lipschitz constant).
However, such conventional approaches are prone to result in suboptimal performance and may cause stability issues if the system conditions change (and the parameters would need to be adapted in real time). 

\subsection{Model-Driven Neural Networks via Deep Unfolding}
In recent years, neural networks (NNs) have been proposed to replace a range of signal processing tasks in communications systems~\cite{OShea2017a,Wang2017a,Mao2018a,Gunduz2019,Qin2019a}. While the performance of such NN-assisted methods is promising in many applications, they suffer from the following drawbacks:
(i)  high computational complexity and memory requirements, and (ii) virtually no performance guarantees are available. 
As an alternative to such black-box methods, \emph{model-driven} NNs are becoming increasingly popular in communications systems~\cite{He2019b}. The idea of model-driven NNs is to fuse principled algorithms that have performance guarantees with tools from NNs, with the goal of combining the best of both worlds. 
\emph{Deep unfolding}~\cite{Hershey2014a} is a powerful instance of such model-driven NNs and is also rapidly gaining popularity in the communications community. 

In the words of the authors of~\cite{Hershey2014a}, deep unfolding can be summarized as follows: ``\emph{[...] given a model-based approach that requires an iterative inference method, we unfold the iterations into a layer-wise structure analogous to a neural network.}''
Put simply, deep unfolding takes an iterative algorithm with a fixed number of iterations $T$, unfolds its structure, and introduces a number of trainable parameters. These parameters are then learned using techniques from deep learning (with suitable loss functions,  stochastic gradient descent, and back-propagation). The resulting unfolded algorithm with learned parameters can then be used to solve a range of tasks in communications systems. 
Deep unfolding has several practical advantages:
(i) Existing performance guarantees for the original iterative algorithms may apply verbatim to learned unfolded networks and appropriate constraints can be imposed on the learned parameters.
(ii) Most unfolded communications algorithms have a relatively small number of trainable parameters, which simplifies training.
(iii) Unfolded algorithms are typically based on well-known methods for which efficient hardware implementations are readily available, which reduces design time. 
(iv) The resulting unfolded algorithms are often intuitive, interpretable, and have low complexity and memory requirements, which is in stark contrast to black-box NNs.

In this survey, we discuss several applications of deep unfolding to communications systems, with a particular focus on MIMO systems and belief-propagation-based decoding of error-correcting codes.\footnote{We note that our formulations of various algorithms may differ slightly from the notation used in the original papers for uniformity.} We also outline other applications and discuss a number of interesting open research directions.

\begin{table*}[t]
	\centering
	\caption{Summary of unfolded learned algorithms for MIMO detection and MIMO precoding.}
	\label{tab:mimo}
	\begin{tabular}{@{}lcccccc@{}}
	\toprule
	Reference				& \cite{Samuel2017a} & \cite{Samuel2019a}	& \cite{Corlay2018a} & \cite{He2018a} & \cite{Takabe2018a} and \cite{Takabe2018b} & \cite{Balatsoukas2019a} \\ 
	\midrule
	Task			& Detection		& Detection		& Detection & Detection & Detection & Precoding \\
	Algorithm		& PGD & PGD	& PGD & OAMP & PGD & PGD \\
	Parameters & $T(12M{\times}2K{+}2K{+}6M)$ 	& $T(12M{\times}2K{+}2K{+}4M)$ & $T(8M{\times}2K{+}2K{+}4M)$ & $2T$ & $2T{+}1$ & $2T$\\
	\bottomrule
	\end{tabular}
\end{table*}

\section{Deep Unfolding for MIMO Systems}\label{sec:mimo}
We now describe applications of deep unfolding to signal processing tasks in multiple-input multiple-output (MIMO) wireless systems. More specifically, we discuss recent results on MIMO detection~\cite{Samuel2017a,Samuel2019a,Corlay2018a,He2018a,Takabe2018a,Takabe2018b,Wei2019a,Khani2019a} and MIMO precoding~\cite{Balatsoukas2019a}, which are also summarized in Table~\ref{tab:mimo}.

\begin{figure*}[t]
	\centering
	\includegraphics[width=0.95\textwidth]{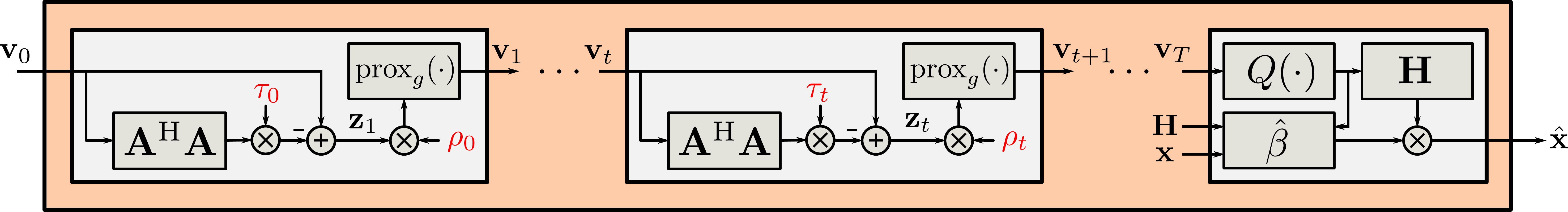}
	\caption{Computation graph structure of the unfolded MU-MIMO 1-bit precoding algorithm used in~\cite{Balatsoukas2019a}, where learnable parameters are highlighted in red. The other MIMO-related work discussed in this section~\cite{Samuel2017a,Samuel2019a,Corlay2018a,He2018a,Takabe2018a,Takabe2018b} build upon very similar unfolded structures.}
	\label{fig:nnoc2po}
\end{figure*}

\subsection{MIMO Data Detection}
The common baseband input-output relation of data transmission over a frequency-flat MIMO channel is as follows:
\begin{align}
	\tilde{\yb}	& = \tilde{\Hb}\tilde{\xb} + \tilde{\nb}. \label{eq:MIMOmodel}
\end{align}
Here, $\tilde{\yb}\in\mathbb{C}^N$ is the complex-valued receive vector, $\tilde{\Hb}\in \mathbb{C}^{N\times M}$ is the $N \times M$ complex-valued MIMO channel matrix, $\tilde{\xb} \in \mathcal{L}^M$ is the vector of transmit symbols, where $ \mathcal{L}$ is the transmit constellation set,  and $\tilde{\nb} \in \mathbb{C}$ is a complex Gaussian noise vector distributed according to $\mathcal{CN}\!\left(0,\sigma ^2\Ib\right)$.

The goal of MIMO data detection is to compute an estimate~$\hat{\xb}$ of the transmitted data vector $\tilde{\xb}$, given the receive vector $\tilde{\yb}$, the channel matrix $\tilde{\Hb}$, and knowledge of the statistics of $\tilde{\nb}$. Maximum likelihood (ML) data detection for this model amounts to solving the following optimization problem:
\begin{align}
	\hat{\xb}	& = \arg\min_{\tilde{\xb} \in \mathcal{L}^M}\|\tilde{\yb}-\tilde{\Hb}\tilde{\xb}\|_2^2. \label{eq:mimodec}
\end{align}
Since the ML data detection problem is NP-hard, computationally efficient approximate methods are used in practice. 
To use NN-based methods for detection, one often operates on real-valued data using the real-valued decomposition of \eqref{eq:MIMOmodel}:
\begin{align}
	\yb	& = \Hb\xb + \nb,  \label{eq:MIMOmodelreal}
\end{align}
where the quantities $\yb \in \mathbb{R}^{2N}$, $\Hb \in \mathbb{R}^{2N\times 2M}$, $\xb \in \mathbb{R}^{2M}$, and $\nb \in \mathbb{R}^{2N}$, are defined as follows:
\begin{align}
	\yb^T & = \begin{bmatrix} \Re\{\tilde{\yb}\}^T &  \Im\{\tilde{\yb}\}^T \end{bmatrix}, &
	\Hb	& = \begin{bmatrix} \Re\{\tilde{\Hb}\} & {-}\Im\{\tilde{\Hb}\} \\ \Im\{\tilde{\Hb}\} & \Re\{\tilde{\Hb}\} \end{bmatrix}\!, \\
	\xb^T & = \begin{bmatrix} \Re\{\tilde{\xb}\}^T &  \Im\{\tilde{\xb}\}^T \end{bmatrix}, & \nb^T & = \begin{bmatrix} \Re\{\tilde{\nb}\}^T &  \Im\{\tilde{\nb}\}^T \end{bmatrix}\!.
\end{align}

The NN-based MIMO data detection methods in~\cite{Samuel2017a,Samuel2019a} are obtained by unfolding the iterations of projected gradient descent data detection algorithms (e.g., the method in \cite{jeon2016performance}) and enriching these algorithms with additional dimensions and trainable parameters. Specifically, in~\cite{Samuel2017a} the following updates are used for $t=0,\ldots,T{-}1$:
\begin{align}
	\zb_t	& =  \text{ReLU} \!\left( \Wb_{1t}\begin{bmatrix} \Hb^T\yb \\ \hat{\xb}_t \\ \Hb^T\Hb\hat{\xb}_t \\ \vb_t \end{bmatrix} + \bb_{1t} \right)\!, \label{eq:Samuel2017a_1} \\
	\hat{\xb}_{t{+}1}	& = \psi_{\kb_t}\! \left(\Wb_{2t}\zb_t + \bb_{2t}\right), \label{eq:Samuel2017a_2} \\
	\hat{\vb}_{t{+}1}	& = \Wb_{3t}\zb_t + \bb_{3t}, \label{eq:Samuel2017a_3}
\end{align}
where $\text{ReLU}(x) = \max\{x,0\}$, $\hat{\xb}_0 = \mathbf{0}$, $\psi_{\kb_t}(\cdot)$ is a soft sign operator parameterized by the trainable parameter vector $\kb_t$, and $\zb_{t}$ is of dimension $2K > 2M$. The quantity $\Hb^T\yb$ is included in the input vector of~\eqref{eq:Samuel2017a_1} because it is a sufficient detection statistic, while $\Hb^T\Hb\hat{\xb}_t$ is an estimate of the (noiseless) sufficient detection statistic at iteration $t$~\cite{Samuel2017a}. The set of trainable parameters is $\left\{\Wb_{1t},\Wb_{2t},\Wb_{3t},\bb_{1t},\bb_{2t},\bb_{3t},\kb_t:t=0,\hdots,T{-}1\right\}$; these parameters can be learned by specifying a suitable loss function and using tools from deep neural networks (such as stochastic gradient descent and back-propagation). 
Finally, hard decisions are obtained by computing $\hat{\xb} = \text{sign}\left(\hat{\xb}_T\right)$. 

The work of~\cite{Samuel2019a} mimics a projected gradient algorithm more closely, thus simplifying the algorithm iteration \eqref{eq:Samuel2017a_1}--\eqref{eq:Samuel2017a_3}:
\begin{align}
	\zb_t	& = \text{ReLU} \! \left( \Wb_{1t}\!\begin{bmatrix}\hat{\xb}_{t} \!-\! \delta_{1t}\Hb^T\yb \!-\! \delta_{2t}\Hb^T\Hb\hat{\xb}_{t} \\ \hat{\vb}_{t} \end{bmatrix} \!+\! \bb_{1t} \right)\!, \\
	\hat{\xb}_{t{+}1}	& = \sigma\!\left(\Wb_{2t}\zb_t \!+\! \bb_{2t}\right), \\
	\hat{\vb}_{t{+}1}	& = \Wb_{3t}\zb_t + \bb_{3t},
\end{align}
where $\sigma(\cdot)$ is a logistic sigmoid, and the trainable parameters $\left\{\Wb_{1t},\Wb_{2t},\Wb_{3t},\bb_{1t},\bb_{2t},\bb_{3t},\delta_{1t},\delta_{2t}:t=0,\hdots,T{-}1\right\}$ are reduced as the $\Wb_{1t}$ matrices are of lower dimension. To aid parameter learning, the cost function in~\cite{Samuel2017a,Samuel2019a} is based on the outputs of all layers and a \emph{residual feature} where the output of each layer is a weighted
average with the output of the previous layer is also added. Details on how to obtain soft outputs and to extend the method to high-order constellations are provided in~\cite{Samuel2019a}. The unfolded data detectors are compared with a wide range of existing detectors in the literature as well as with a standard NN-based detector over different channel models and is shown to provide competitive performance.

To enable support for higher-order constellations, the work of \cite{Corlay2018a} proposes to use a sum of shifted sigmoid functions:
\begin{align}
	\psi(x)	& = \textstyle \sum_{l=1}^L\sigma (x-\tau_i) + A.
\end{align}
Here, the shifts $\tau_i$ are pre-defined based on the constellation set $\mathcal{L}$ and $A$ is a fixed offset. Moreover, two distinct NN-based detectors are trained with different initialization strategies and the best output is kept. It is also argued that only ML-detectable training samples should be used for training. Simulation results show that close-to-ML performance is achieved for a constellation with $L=5$ levels over a fixed (during training and data detection) MIMO channel with $N=M=8$.

The work of~\cite{He2018a} unfolds the iterations of an orthogonal approximate message passing (OAMP) detector~\cite{jeon2016performance}, where only trainable scalars $\left\{\gamma_t,\theta_t:t=0,\hdots,T{-}1\right\}$ are introduced. The following updates are used for $t = 0,\hdots,T{-}1$:
\begin{align}
	\rb_t	& = \hat{\xb}_t + \gamma_t \Wb_t(\yb-\Hb\hat{\xb}_t), \\
	\hat{\xb}_{t{+}1} & = \mathbb{E}\left[\xb \mid \rb_t,\tau_t\right], \\
	v_t^2	& = \textstyle \frac{\|\yb-\Hb\hat{\xb}_t\|_2^2 - M\sigma^2}{\text{tr}(\Hb^T\Hb)}, \\
	\tau_t^2 & = \textstyle \frac{1}{2N}\mathrm{tr}(\Cb_t\Cb_t^T)v_t^2 + \frac{\theta_t^2\sigma^2}{4N}\mathrm{tr}(\Wb_t\Wb_t^T),
\end{align}
where $\Wb_t$ is a function of the channel matrix $\Hb$, $v_t^2$, and~$\sigma^2$, and $\Cb_t$ is a function of $\Hb$ and $\Wb_t$, as defined in~\cite{He2018a}. Simulation results for Rayleigh and correlated channels demonstrate that data-driven tuning of $\gamma_t$ and $\theta_t$ can lead to significant performance improvements compared to standard OAMP.

The algorithms in~\cite{Takabe2018a,Takabe2018b} target detection for massive overloaded MIMO channels, i.e., channels where $N \gg M$. The proposed data detection algorithm is based on projected gradient descent for a total of $T$ iterations and with the introduced trainable scalars $\left\{\alpha,\gamma_t,\theta_t:t=0,\hdots,T{-}1\right\}$:
\begin{align}
	\rb_t	& = \hat{\xb}_t + \gamma_t\Wb(\yb-\Hb\hat{\xb}_t), \\
	\hat{\xb}_{t{+}1}	& = \tanh\! \left({\rb_t}/{|\theta_t|}\right),
\end{align}
where $\hat{\xb}_0 = \mathbf{0}$ and $\Wb = \Hb^T(\Hb\Hb^T+\alpha \mathbf{I})^{{-}1}$. The authors use incremental training to avoid vanishing gradient problems. Simulation results show that the trained projected gradient detector provides similar performance to other detectors but at a significantly lower complexity. Moreover, the trained projected gradient detector is also shown to perform well in traditional, i.e., non-overloaded, MIMO systems.

Very recently, the papers \cite{Wei2019a} and \cite{Khani2019a} used deep unfolding for MIMO data detection based on conjugate gradients and projected gradient descent, respectively; both methods achieve near-ML performance at low complexity.

\subsection{Multi-User (MU) MIMO Precoding}
MU-MIMO precoding consists of multiplying the transmit symbol vector $\tilde{\xb}$ with a precoding matrix $\Pb$ so that a suitably defined performance metric (e.g., the SNR at the receiver) is maximized. The complex-valued system model in~\eqref{eq:MIMOmodel} is
\begin{align}
	\tilde{\yb}	& = \tilde{\Hb}\tilde{\Pb}\tilde{\xb} + \tilde{\nb} = \tilde{\Hb}\tilde{\vb} + \tilde{\nb},
\end{align}
where $\tilde{\vb} = \tilde{\Pb}\tilde{\xb}$ and the corresponding real-valued model of~\eqref{eq:MIMOmodelreal} using $\yb$, $\Hb$, $\vb$, and $\nb$ can be derived accordingly.

The results in~\cite{Balatsoukas2019a} describe a projected gradient descent algorithm for precoding in massive MU-MIMO systems with $1$-bit quantization at the transmitter. In this scenario, each element of the vector $\vb$ is constrained to the binary set $\{{-}\upsilon,{+}\upsilon\}$, where $\upsilon^2 = {\frac{P}{2M}}$ is selected to satisfy a transmit power constraint $P$. 
The following updates with trainable scalars $\left\{\tau_t,\rho_t:t=0,\hdots,T{-}1\right\}$ are used for a total of $T$ iterations:
\begin{align}
	\zb_{t{+}1} &= \vb_{t{+}1} - \tau_t \Ab^H\Ab \vb_{t}, \label{eq:nnoc2po_z} \\
	\vb_{t{+}1} & = \mathrm{prox}_g ( \zb_{t{+}1};\rho_t,\xi) \label{eq:nnoc2po_v}.
\end{align}
where  $\mathrm{prox}_g ( \zb;\rho_t,\xi) = \mathrm{clip}\!\left(\rho\Re\{\zb\}, \xi\right) + j\mathrm{clip}\!\left(\rho\Im\{\zb\}, \xi \right)$
is the \emph{proximal operator}, $\Ab = \left(\Ib-\xb\xb^T/\|\xb\|_2^2\right)\Hb$, and $\vb_T$ is quantized to $\{{-}\upsilon,{+}\upsilon\}$. Simulation results for a range of channel models show that learning suitable parameters $\tau_t, \rho_t$ allows one to decrease the number of iterations $T$ by a factor of two for the same error-rate performance. The computational graph corresponding to the unfolded version of~\eqref{eq:nnoc2po_z} and \eqref{eq:nnoc2po_v} along with the final quantization $Q(\cdot)$ and transmission over~$\Hb$ is shown in Fig.~\ref{fig:nnoc2po}.

\section{Deep Unfolding for Belief-Propagation-Based Channel Decoding}\label{sec:bp}
Belief propagation is an iterative message-passing algorithm that is commonly used to decode error-correcting codes. 
The message-passing strategy is typically described by a bipartite Tanner graph that represents the parity-check matrix of the code. These Tanner graphs consist of two types of nodes, namely \emph{variable nodes} and \emph{check nodes}. Each variable node is associated with a codeword bit and each check node is associated with a parity-check equation. 
Let $\mathcal{V}$ denote the set of variable nodes, $\mathcal{C}$ denote the set of check nodes, and $\mathcal{N}(x)$ denote the set of (one-hop) neighbors of a node $x$. Then, for each $v \in \mathcal{V}, c \in \mathcal{N}(v)$, the variable-to-check messages $m_{t}^{v\rightarrow{c}}$ at iteration $t \in \{1,\hdots,T\}$ are:
\begin{align}
	m_{t}^{v\rightarrow{c}}	& = \textstyle  l^v + \sum _{c' \in \mathcal{N}(v)\setminus{c}} m_{t{-}1}^{c'\rightarrow{v}}, \label{eq:bpv}
\end{align}
where $l^v$ denotes the channel log-likelihood ratio (LLR) for variable node $v$ and $m_{0}^{c\rightarrow{v}} = 0$ by convention. Moreover, for each $c \in \mathcal{C}, v \in \mathcal{N}(c)$, the check-to-variable messages $m_{t}^{c\rightarrow{v}}$ at iteration $t$ are given by
\begin{align}
	m_{t}^{c\rightarrow{v}}	& = \textstyle  2\tanh^{{-}1}\!\left( \prod_{v' \in \mathcal{N}(c)\setminus{v}} \tanh \left(\frac{m_{t}^{v\rightarrow{c}}}{2} \right) \right)\!. \label{eq:bpc}
\end{align}
For each $v \in \mathcal{V}$, the bit-decision metric is calculated as
\begin{align}
	m_{t}^{v}	& = \textstyle  l^v + \sum _{c \in \mathcal{N}(v)} m_{t{-}1}^{c\rightarrow{v}}, \label{eq:bpd}
\end{align}
and final bit-decisions are generated as follows:
\begin{align}
	\hat{u}_t^v	& =\textstyle  \frac{1}{2}\left(1- \text{sign}(m_{t}^{v})\right). \label{eq:bphard}
\end{align}

In~\cite{Nachmani2016a}, the variable-to-check BP equation in \eqref{eq:bpv} is modified by adding trainable weights $w_t^{v}$ and $w_t^{c'}$, which yields:
\begin{align}
	m_{t}^{v\rightarrow{c}}	& \textstyle = w_t^{v}l^v + \sum _{c' \in \mathcal{N}(v)\setminus{c}} w_{t{-}1}^{c'}m_{t{-}1}^{c'\rightarrow{v}}. \label{eq:bpvweights}
\end{align}
Moreover, bit-decisions are generated using the following soft (i.e., differentiable) version of \eqref{eq:bphard}:
\begin{align}
	\hat{u}_t^v	& = \sigma\! \left(m_{t}^{v}\right).
\end{align}
The authors use a binary cross-entropy loss function that uses the $\hat{u}_t^v$ values from all iterations $t$ in order to aid learning and avoid vanishing gradient problems. The unfolded BP decoder is trained using synthetically-generated training data for a range of different signal-to-noise-ratio (SNR) values. Simulation results for a variety of  BCH codes show that the unfolded BP decoder with learned weights significantly outperforms traditional BP decoders.

The methods in~\cite{Nachmani2017a,Nachmani2018a} improve upon~\cite{Nachmani2016a} by using a recurrent neural network (RNN) structure so that the weights $w_t^{v}$ and $w_t^{c'}$ do not change over the iterations. Moreover, the authors use a technique called \emph{relaxation} where consecutive messages are combined using learned weights, which improves the convergence behavior of the BP decoder. Moreover,~\cite{Lugosch2017a,Nachmani2018a} simplify~\cite{Nachmani2016a} by using normalized min-sum (MS) decoding for the check nodes with a learned parameter~$w$:
\begin{align}
	m_{t}^{c\rightarrow{v}}	 = \, &\,  \textstyle w \times \prod_{v' \in \mathcal{N}(c)\setminus{v}} \text{sign} \left(m_{t}^{v\rightarrow{c}}\right) \notag \\
	& \times \min_{v' \in \mathcal{N}(c)\setminus{v}} |m_{t}^{v\rightarrow{c}}|. \label{eq:msc}
\end{align}

The method in~\cite{Xu2017a} uses an unfolded normalized  MS algorithm for the decoding of polar codes. The main difference with~\cite{Lugosch2017a,Nachmani2018a}, apart from the slightly different message scheduling required to decode polar codes, is that the normalization parameter $w$ is allowed to differ for every message and for every iteration. Simulation results for polar codes of various block-lengths and rate $R=1/2$ show that the unfolded MS decoder with per-message learned normalization parameters outperforms the standard normalized MS decoder by approximately $0.5$~dB. The authors also provide a high-level discussion of hardware implementation considerations.

The authors of~\cite{Cammerer2017a} propose a hybrid BP-NN decoder for polar codes, where a fraction of the messages is calculated using standard BP message-passing rules, while the remaining messages are calculated using trained NNs. This approach enables the scaling of NN-assisted decoders to large block-lengths, while simulation results show very competitive performance with respect to conventional decoders for polar codes. 

The method in~\cite{Xu2018a} unfolds the MS algorithm to decode polar codes. The authors first use a method to convert the message-passing graph of polar codes into a conventional sparse Tanner graph so that the standard BP message-passing rules of~\eqref{eq:bpv} and \eqref{eq:bpc} can be used verbatim, thus avoiding the different message schedule used in~\cite{Xu2017a,Cammerer2017a}. Moreover, a single weight $w'$ is used for all variable-to-check messages at all iterations so that~\eqref{eq:bpvweights} is simplified to:
\begin{align}
	m_{t}^{v\rightarrow{c}}	& = \textstyle  l^v + \sum _{c' \in \mathcal{N}(v)\setminus{c}} w' m_{t{-}1}^{c'\rightarrow{v}}.
\end{align}
The non-normalized MS update rule is used for the check-to-variable messages, i.e., \eqref{eq:msc} with $w=1$. Simulation results show that the use of a single weight $w'$ has a negligible effect on the error rate of the decoder, while significantly reducing the complexity of both learning and decoding.

The papers~\cite{Wu2018a,Dai2018a} propose to unfold the normalized-offset MS algorithm to decode LDPC codes and polar codes, respectively. The minimum-finding part in~\eqref{eq:msc} is replaced by:
\begin{align}
	\alpha_t^{v\rightarrow{c}} \cdot \min_{v' \in \mathcal{N}(c)\setminus{v}} \max\left(|m_{t}^{v\rightarrow{c}}|-\beta_t^{v\rightarrow{c}},0\right),
\end{align}
where $\alpha_t^{v\rightarrow{c}}$ and $\beta_t^{v\rightarrow{c}}$ are per-message and per-iteration trainable parameters. Simulation results show that these additional parameters can improve the performance of unfolded MS decoding with respect to standard MS and BP decoding as well as previous works on unfolded MS decoding.

In~\cite{Doan2018b}, a joint CRC-polar MS decoding algorithm is proposed, which exploits the concatenated factor graph of a polar code and a CRC. Similarly to previously described works, trainable weights are assigned to the edges of the unfolded factor graph. Moreover, a multi-loss cost function is used to improve the training process. In general, multi-loss functions are a sum of multiple loss functions from different parts of the NN to be trained. In this case, the multi-loss function takes the outputs of both the MS part and the CRC part of the factor graph into account. Simulation results show improved performance with respect to~\cite{Nachmani2018a,Xu2017a}.

The method in~\cite{Lian2019a} uses an unfolded structure that resembles that of~\cite{Nachmani2018a}, with the main difference that a single weight is used for all messages and all iterations. The authors also argue that the binary cross-entropy function that is commonly minimized to train unfolded decoders does not necessarily minimize the bit error rate (BER). Instead, they propose a new cost function that is based on the so-called \emph{soft bit error} concept. For a single bit, if the actual bit-value is $a \in \{0,1\}$ and the estimated (soft) bit-value at the output of the unfolded decoder is $b \in [0,1]$, the soft bit error $L_{\text{sbe}}(a,b)$ is given by:
\begin{align}
	L_{\text{sbe}}(a,b)	& = (1-b)^ab^{1-a},
\end{align}
whereas the standard binary cross-entropy $L_{\text{bce}}(a,b)$ would be:
\begin{align}
	L_{\text{bce}}(a,b)	& = -\log\left(b^a(1-b)^{1-a}\right).
\end{align}
Instead of training for a single SNR or a set of SNR points, the authors of~\cite{Lian2019a} use an auxiliary NN that learns parameter values given the SNR as an input.

Finally, the work of \cite{Vasic2018a} proposes the idea of unfolding in order to learn finite-alphabet (FA) decoding of LDPC codes. In FA decoding, messages are quantized using a very small number of quantization bits and it is thus crucial that the quantization thresholds and levels are designed very carefully. The authors show that by unfolding and learning FA decoders, gains of up to $0.25$~dB can be achieved for a $(1296,972)$ QC-LDPC code when using $3$ quantization bits.

\section{Deep Unfolding for Other Communications~Applications}\label{sec:other}
There exist a plethora of other communications applications in which the idea of unfolding has been used---we now briefly summarize some of these applications. For channel decoding that is not based on BP decoding, references~\cite{Kim2018a,Jiang2019a,He2019a} study unfolding of Turbo decoding, whereas~\cite{Doan2018a} discusses successive cancellation decoding of polar codes. The work in~\cite{Shlezinger2019a} proposes to replace the channel-dependent parts of the Viterbi detection algorithm by a DNN. 
NNs have also been used extensive for non-linear signal processing tasks. In this case, unfolding does not refer to the iterations of some algorithm, but rather to the non-linear equations themselves (e.g., parallel Hammerstein model~\cite{Janczak2003a,Yu2014a} or Schr\"odinger wave equation). This approach has been recently applied to optical communications (e.g., \cite{Hager2018a,Hager2018b}) and to full-duplex communications (e.g., \cite{Balatsoukas2018a}). Finally, unfolding has been extensively applied to the iterative shrinkage-thresholding algorithm (ISTA) to solve sparse linear inverse problems~\cite{Gregor2010a,Borgerding2016a,Kamilov2016a,Borgerding2017a,Ito2019a,Takabe2019a}, which is a general tool that finds use in communications systems (e.g., for sparse channel estimation).

\section{Future Research Directions}\label{sec:future}
Even though the idea of deep unfolding is relatively novel and many open research questions remain, it has already found wide applicability in communication systems and is likely to transform a range of other signal processing tasks. For example, proximal algorithms~\cite{Parikh2014a,goldstein2014field} solve a wide range of optimization problems in communication systems and they are generally well-suited for unfolding.  
We conclude this brief survey by outlining some interesting future research directions.

\subsection{Unfolded Structures with Acceleration Methods}
Several techniques that have been used in the optimization literature to accelerate the convergence of optimization algorithms can be incorporated into unfolded architectures with trainable parameters. Some examples of these techniques include preconditioning, momentum and  Onsager terms, restart, and adaptive step-size rules. Such methods are particularly interesting for severely iteration-constrained applications where obtaining the fastest possible convergence is of the utmost importance. It may also be beneficial, from both a complexity and performance perspective, to derive and optimize unfolded structures  that directly operate on complex-valued signals.

\subsection{Loss Functions} 
Novel application-tailored cost functions can improve the convergence of the training process. This can not only lead to better results for the same computational effort, but it may also enable real-time and online training of unfolded structures. Some examples of customized loss functions already exist (e.g., \cite{Lian2019a} uses a soft bit error function, \cite{Lugosch2018a} proposes a syndrome-based cost function, and~\cite{Liu2019a} uses a cost function that is tailored to the quantum error-correction scenario), but they are mostly limited to channel decoding. All works on MIMO detection/precoding that we have described~\cite{Samuel2017a,Samuel2019a,Corlay2018a,He2018a,Takabe2018a,Takabe2018b,Balatsoukas2019a} use the standard mean-squared error (MSE) cost function. The MSE cost function has the advantage that closed-form solutions or very accurate iterative approximations can be derived in many cases. However, the MSE is not necessarily a good proxy for the error rate performance of the system. For example, in a multi-user MIMO setting, the system error rate will most likely be dominated by the user with the largest MSE. When optimizing an unfolded algorithm, even when the algorithm itself has been derived assuming an MSE cost function, it is typically easy to learn the optimal set of parameters for a different cost function. In the multi-user MIMO example mentioned previously, this could be the maximum MSE over all users, or even the maximum MSE over all users and all channel realizations in the training dataset. 

\subsection{Training} 
Even though in many applications training can be carried out offline, it is still a task that requires considerable effort and thus deserves attention. In applications where it is sufficient to unroll a small number of iterations, training is mostly straightforward. However, when more iterations are considered, numerous problems arise. A common problem is that of vanishing gradients, where it becomes increasingly difficult to find suitable parameters in early iterations. This problem can be addressed by using multi-loss functions (like most of the works presented in this survey), incremental training~\cite{Takabe2018a}, or by simply using a set of known good initial values to minimize training~\cite{Balatsoukas2019a}.
Another solution would be to perform windowed training, where unfolded iterations are trained only over a moving window of fixed size. This approach can also significantly reduce the memory required for training, which may become a limiting factor when considering algorithms with high-dimensional inputs (e.g., in massive MIMO) and a large number of iterations, since all intermediate output values of each mini-batch need to be stored for back-propagation. Online training methods that adapt to, e.g., changing channel or SNR conditions, is another important problem. Finally, it is often unclear what the best dataset for training is. We note that some preliminary works already focused specifically on this direction~\cite{Benammar2018a,Beery2019a}, but more research is required.
 
\subsection{Hardware Implementation}
Unfolded learned algorithms are particularly attractive from a hardware implementation perspective, as they strongly resemble known algorithms for which efficient hardware architectures already exist. However, the hardware implementation complexity aspect is typically not considered in the literature, with some notable exceptions being~\cite{Xu2017a}, where high-level hardware considerations for unfolded MS decoding are discussed, and~\cite{Balatsoukas2018b}, where FPGA and ASIC implementation results of the method in~\cite{Balatsoukas2018a} are presented. As such, it remains largely unclear how the additional trainable parameters required by unfolded algorithms affect the hardware implementation complexity and the achieved throughput. Moreover, efficient hardware implementations of the training step are necessary for situations that require online learning.

\balance
\bibliographystyle{IEEEtran}
\bibliography{19SiPS_CommsUnfolding}
\balance

\end{document}